\documentclass[11pt]{article}
\newif\ifptep
\ptepfalse

\usepackage{graphicx}
\usepackage{hyperref}

\begin{document}

\title{
Accelerator Control System at KEKB and the Linac
}
\ifptep 

\author{
\name{Kazuro Furukawa}{1,\ast},
\name{Atsuyoshi Akiyama}{1}, 
\name{Eiichi Kadokura}{1},
\name{Miho Kurashina}{1},
\name{Katsuhiko Mikawa}{1},
\name{Tatsuro Nakamura}{1},
\name{Jun-ichi Odagiri}{1},
\name{Masanori Satoh}{1}, and
\name{Tsuyoshi Suwada}{1}
}

\address{\affil{1}{High Energy Accelerator Research Organization (KEK), Oho 1-1, Tsukuba, Ibaraki, 305-0801, Japan}
\email{kazuro.furukawa@kek.jp}}

\else 

\author{
Kazuro Furukawa\thanks{\texttt{<\,kazuro.furukawa\,@\,kek.jp\,>}}, 
Atsuyoshi Akiyama, 
Eiichi Kadokura, \\
Miho Kurashina, Katsuhiko Mikawa, Tatsuro Nakamura, \\
Jun-ichi Odagiri, Masanori Satoh, Tsuyoshi Suwada\\
KEK, Tsukuba, Ibaraki, 305-0801, Japan
}
\date{}
\newcommand\etal{\textit{et~al.}}
\maketitle

\fi


\begin{abstract}

KEKB has completed all of the technical milestones and has offered important insights into the flavor structure of elementary particles, especially CP violation. The accelerator control system at KEKB and the injector linac was initiated by a combination of scripting languages at the operation layer and EPICS (experimental physics and industrial control system) at the equipment layer. 
During the project, many features were implemented to achieve extreme performance from the machine. In particular, the online linkage to the accelerator simulation played an essential role.  In order to further improve the reliability and flexibility, two major concepts were additionally introduced  later in the project, namely, channel access everywhere and dual-tier controls. 
Based on the improved control system, a virtual accelerator concept was realized, allowing the single injector linac to serve as three separate injectors to KEKB's high-energy ring, low-energy ring, and Photon Factory, respectively.  These control technologies are indispensable for future particle accelerators. 
\end{abstract}

\ifptep 

\subjectindex{accelerator controls, electron-positron collider}

\maketitle

\fi

\section{Introduction}

The KEKB B-Factory was designed as an asymmetric electron/positron collider in order to study the violation of CP symmetry in the B-meson system.  It consisted of double storage rings of an 8~GeV electron high-energy ring (HER) and a 3.5~GeV positron low-energy ring (LER) with a diameter of 1~km, and a full energy injector linac of 600~m~\cite{ptep-kekb}.  It achieved a collision luminosity of $2.11\times10^{34}\,{\rm cm}^{-2}\,{\rm s}^{-1}$, twice as much as the designed luminosity, and led to Kobayashi and Maskawa winning the Nobel Prize for the theory of quarks and CP symmetry violation.  

During a decade of successful operation, KEKB completed all of the technical milestones and offered important insights into the flavor structure of elementary particles. 

KEKB collider and Belle detector were constructed and operated at almost the same time as PEP-II and Babar at SLAC with the same scientific goal.  This provided a friendly competition between these two projects and led to many collaborative efforts~\cite{ptp-kekb}.  

\section{KEKB accelerator control system}

The success of the high performance operation of KEKB owed much to the control system. It was designed more than 15 years ago and started beam operation in 1998.  While it inherited some of its resources from the previous project, TRISTAN, it restructured most of the software and hardware components.  It employed the EPICS (experimental physics and industrial control system) toolkit for the low-level control mechanism and scripting languages for high-level operational applications.  This combination provided a flexible and robust operational environment.  The details of the hardware and software architecture at the commissioning of KEKB are described elsewhere~\cite{nim-bcont}. 

\subsection{Lower-level controls with EPICS}

Before KEKB, projects in the institute repeated the development of their own control systems.  As technologies such as computers, system software, and field networks
became de facto standards at the beginning of the 1990s, the sharing of control system architectures between projects was considered.  After SSC (superconducting super collider) chose EPICS as its main control toolkit, EPICS became a candidate for future controls at KEK~\cite{epics}.  EPICS was studied at the injector linac, and EPICS gateways were constructed based on the existing control system~\cite{ical95-epics}.  In contrast, for the KEKB ring it was decided to employ EPICS controls and the previous software resources were not employed.  The main reason for this was that the linac already had a network-based control system and it had to continue operation for light sources even during the upgrade construction towards KEKB, but the ring could have shut down the accelerator completely for 4 years.  

The KEKB ring employed several fieldbuses such as VME, VXI, CAMAC, GPIB, and ARCNET depending on the intended purpose.  Approximately 100 VME systems with the VxWorks operating system served as EPICS input/output controllers (IOCs) for all the hardware devices, including 200 VXI mainframes, 50 CAMAC crates, and 200 ARCNET segments.  

For the networks and computers in the global control system, a quite standard environment was applied, taking account of reliability, such as redundant GbE network systems, redundant network file servers, blade CPU servers, and commodity console machines with X11 server software at KEKB and the linac. 

At the linac most of the devices employed controllers with Ethernet and TCP/IP instead of other kinds of fieldbuses.  Before the KEKB project, network-based PLCs (programmable logic controllers), CAMAC crate controllers, and VME computers were managed by middle-layer software on Unix servers.  During the upgrade construction for KEKB, these network-based controllers were shared between the old control software and the EPICS gateways.  Gateways were implemented with a portable channel access server (PCAS) at the beginning, and were eventually replaced by soft IOCs as EPICS started to support Unix-based IOCs~\cite{ical07-bcont}.  

The number of EPICS process variables in the system was approximately 300,000, and that of hard-disk-archived ones was 150,000.  They were distributed over 150 VME-based and Linux-based EPICS IOCs. 

\subsection{High-level application with scripting languages}

At the linac, the Tcl/Tk scripting language was effectively employed for its commissioning~\cite{ical99-linac} after the language had been utilized for testing tools for many years.  Later, for both the ring and the linac, Python was employed as it had more strong points~\cite{ical07-python}.  Much of the device manipulation software was written in these two languages, as well as MEDM.  

For beam operation, SAD (strategic accelerator design program) was extended to have an interpreter, SADscript, which emulated most of the Mathematica language~\cite{apac98-sad, sadweb}.  This provided most of the functionalities required for the accelerator operation, such as linear beam optics, symplectic beam tracking, non-linear beam dynamics, optimization, list processing, numerical manipulation, EPICS channel access, and a graphical user interface.  Thus, online linkage between the beam simulation and the machine operation was provided through the SAD and EPICS environments. 

In fact, many slow closed feedback loops were implemented using these script languages in order to stabilize the beam characteristics and to maintain the ultimate collision condition.  Script languages are very suitable for rapid prototyping.  If a program needed higher performance, the algorithm was eventually transported into a faster IOC process.  Such feedback loops were often effective in suppressing interim instability of hardware devices before repair. They were also important during the beam studies because the beam conditions were very different from the normal ones.  While a certain parameter was scanned in such studies, other parameters often had to be maintained stable. 

During normal operation it is necessary to measure the beam response on certain parameter changes, and then to optimize those parameters.  Such a process can be interactively carried by SADscript, and then turned into a graphical user interface that is performed routinely.  New ideas for luminosity optimization were often proposed in the morning meeting, and corresponding operational tools were realized within a day or two.  Some of the ideas turned out to be favorable, and the tool became utilized routinely.  As many equipment and beam-physics ideas were proposed day by day, rapid tool development was crucial and SADscript played a significant role.  Actually, it is difficult to name a single mechanism that enabled the KEKB's high performance; however, the accumulation of a hundred 1\% improvements provided twice the performance.  

Besides script-based application programs, many software packages were developed, including a data archiver (KEKBlog), an alarm handler (KEKBalarm), an electric operational logbook (Zlog), an archive browser, etc. The combination of them refined the everyday KEKB operation. 


\section{Two additional concepts towards higher luminosity}

As an upgrade to a control system needs considerable effort, it is preferable to maintain the same environment during a project.  However, an accelerator project can span more than ten years, and related software, hardware, persons, companies, and their policies may change substantially during that period.  Thus, it is necessary to introduce advanced technologies to improve the machine even during the project.  On the other hand, it is often difficult to accommodate new technologies with existing components.  If a component is modified to accept them, others may have to be modified.  

Actually, during the KEKB project, new operation schemes were introduced almost every year.  As most components in the control system were kept the same, it was rather difficult to catch up with the requirements at a later period.  
One modification might trigger another, and several modifications had to be performed at the same time.  Because the shutdown period in a year was limited, if the extent of modifications exceeded the limit, it became very hard.  

Thus, we needed to be prepared to accommodate small upgrades each year not only for application programs but also for control system infrastructure, including base software and hardware components, in order to manage a project for a long period. 


Without changing the architecture of the control system, featured with EPICS and scripting languages, new concepts were incorporated at later years in the project.  These included channel access everywhere and the dual-tier control system, described below.  

\subsection{Channel access everywhere}

The accelerator control architecture in KEK evolved in several steps.  Some time ago some control systems were standardized with a combination of several fieldbuses, VME field computers, and Unix computers.  In order to consolidate the efforts on the development and maintenance, some of the fieldbuses were gradually removed and many controllers were directly attached to IP networks.  

\begin{figure}[tb]
   \centering
  \includegraphics*[width=100mm]{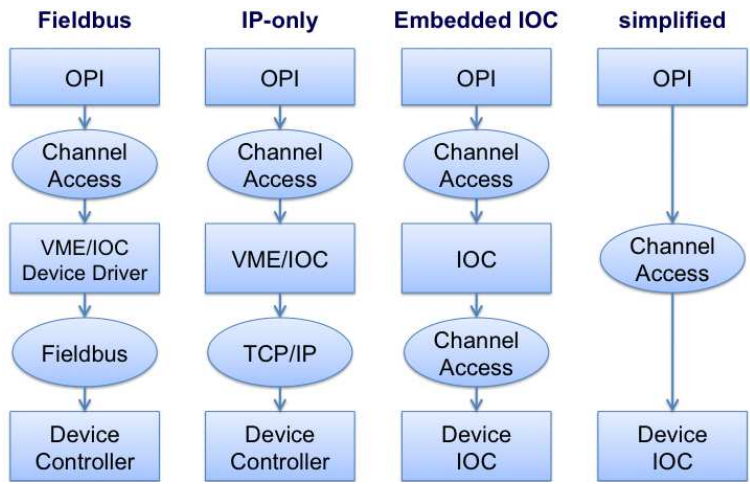}
   \caption{Evolution of device controllers (from left to right) from fieldbus devices towards CA everywhere with embedded IOC.}
   \label{devcont}
\end{figure}

At the same time, the EPICS control software framework was employed at control systems at KEK.  Eventually, many controllers evolved to embed the same EPICS IOC software as on VME field computers, as illustrated in Fig.~\ref{devcont}.  Common pieces of IOC software communicated with others using a common protocol called channel access (CA), which realized a unified application development environment from top to bottom.  We call this embedded EPICS framework ``CA everywhere'', and it enabled both rapid development and smooth maintenance~\cite{ical07-bcont, pac07-kekcont}.  

Different kinds of controllers have been developed in the framework of CA everywhere, and they greatly reduced the control efforts and improved the reliability. They include the following controllers.  Many of them were realized employing PLC~\cite{ical09-plc}, embedded computers~\cite{ical09-bpm}, or FPGA (field programmable gate array)~\cite{ical11-llrfcont}. 
\begin{itemize}
\item Temperature monitor with Linux
\item Yokogawa FAM3 PLC with realtime Linux
\item Tektronix oscilloscope for 50~Hz measurement with Windows
\item Time-to-digital converter (TDC) with FPGA and Linux
\item MicroTCA LLRF controller with FPGA and Linux
\item Microwave power modulator with FPGA and Linux
\item Libera BPM readout at 50~Hz with FPGA and realtime Linux
\item NI Compact-RIO with CAS and FPGA
\end{itemize}

\subsection{Dual-tier control system}

For higher experimental performance at KEKB and the light sources that share the same injector, it was favorable to inject beams in top-up mode into all the storage rings. In the Photon Factory (PF), a stable beam current would provide precise experimental results. In KEKB, stability was desired for sensitive beam collision tuning to increase the luminosity. 

To that end, simultaneous top-up injection had been established for three storage rings at the KEKB HER, LER, and PF since 2009. Electron and positron beams with very different characteristics, charged from 0.1~nC to 8~nC and with energies between 2.5~GeV and 8~GeV, were exchanged at a rate of 50~Hz (20~ms). As a result, stored beam stabilities of 0.05\% (1~mA) at the KEKB HER and LER, and 0.01\% (0.05~mA) at PF were achieved, improving the quality of experiments.  

While it initially took 30 seconds to 2 minutes to switch the beam injection modes between these storage rings, it was preferable to change many parameters synchronously within 20~ms.  Thus, global and fast controls were established for such pulse-to-pulse beam modulation. This kind of beam modulation was realized a long time ago but the speed was much slower, around one to several seconds~\cite{pac77-ppm}. As the existing control system, based on ten-year-old hardware and conventional EPICS software, was inadequate for controlling the beam within 20~ms, a new control system with an event notification mechanism, capable of regulating $\sim$150 parameters at 50~Hz, was installed. This event-based control system covered the controls of the low-level RF, high-power RF, pulsed magnets, an electron gun, injection systems, and beam instrumentation, whose devices were spread over a distance of 1~km. While the event-based control system was supervised by the EPICS control software, it had a dedicated communication link for fast, global, and robust controls~\cite{ical09-event}. 

\begin{figure}[tb]
   \centering
   \includegraphics*[angle=0, width=120mm]{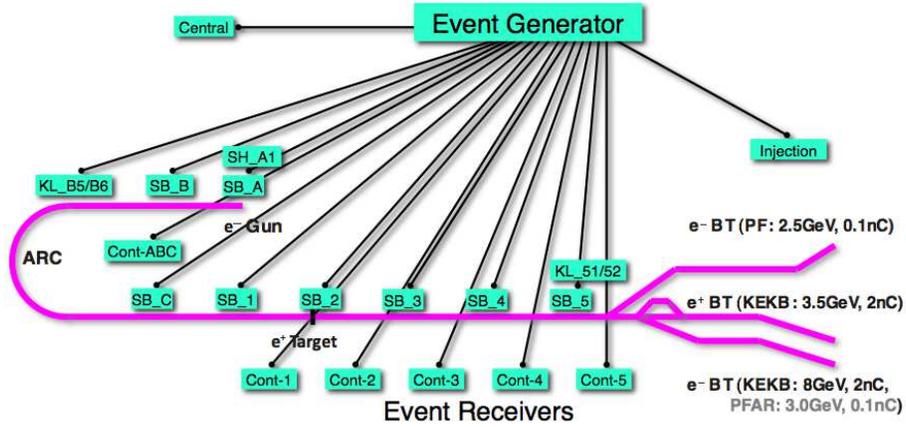}
   \caption{Overall configuration of the event-based control system at the injector linac.  17 event receiver stations cover the 1 km facility.}
   \label{fig-config}
\end{figure}

The event generator sent timing signals and control data to 17 event receiver stations arranged in a star-like topology as in Fig.~\ref{fig-config}. Each link between the event generator and a receiver was carried over a single optical fiber. It provided both synchronized timing signals, with approximately 10~ps precision, and synchronized controls through a realtime software mechanism, at about 10~{$\mu$}s precision. Recent technological advances in FPGA and SFP (small form-factor pluggable) enabled reliable controls in this configuration.  

VME-based event control modules of the generator (EVG230) and the receiver (EVR230RF) from MRF were utilized~\cite{mrfweb}. The event generator provided several events corresponding to beam and device controls synchronized to one of the linac RF clock (114.24~MHz). 

\begin{figure}[tb]
   \centering
  \includegraphics*[width=100mm]{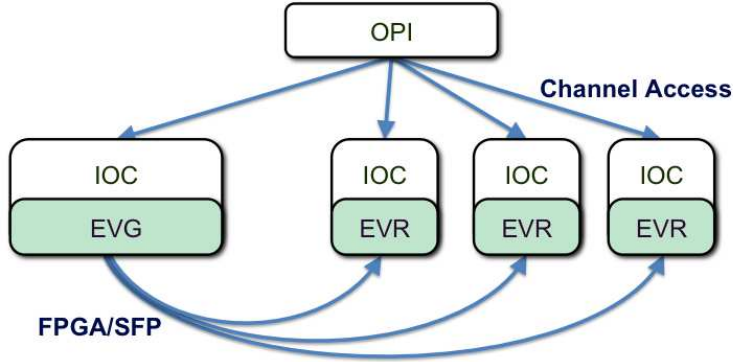}
   \caption{Dual-tier controls with EPICS channel access at the top and fast event synchronized control at the bottom.}
   \label{duallayer}
\end{figure}

The same dual-tier control system with a conventional EPICS control and an event-based control depicted in Fig.~\ref{duallayer} will also be essential in the future SuperKEKB. Simultaneous injection will be maintained, as the beam lifetime will be more limited at the SuperKEKB HER and LER. Many more parameters will have to be managed precisely in order to realize lower-emittance beams for higher luminosity.  

The event-based control tier manages global and fast controls in the pico- to microsecond range. The EPICS control tier covers slower parameter controls for the event-based controls 
as well as existing conventional controls. The EPICS tier arbitrates operation requests of average beam repetition rates from the three rings, and schedules different beams pulse by pulse.\footnote{Under typical operating conditions, the average injection rates were 25~Hz for LER, 12.5~Hz for HER and 0.5~Hz for PF.  However, they were frequently changed to maintain the stored beam current.}  Such requests occurred every several seconds, and upon each request the beam mode schedule was reprogrammed at the event-control tier through EPICS CA. 
Such a dual-tier control system is also optimal for the next generation of accelerator systems. 

\section{Virtual accelerators}

Under the simultaneous injection configuration, the event-based control system provided beam-mode dependent control parameters.  Moreover, these parameters in different beam modes were organized to be independent both for controls and measurements.  Thus, we can see these independent parameter sets as independent virtual accelerators.  For each 20 ms time slot, the event system associated one of the virtual accelerators with the real accelerator. 

Because these control parameters for each virtual accelerator continue to exist, human operators and operational software can act on one of these virtual accelerators without any interference from the others. 

\subsection{Event-based beam feedback loops}

\begin{figure}[tb]
   \centering
   \includegraphics*[angle=0, width=100mm]{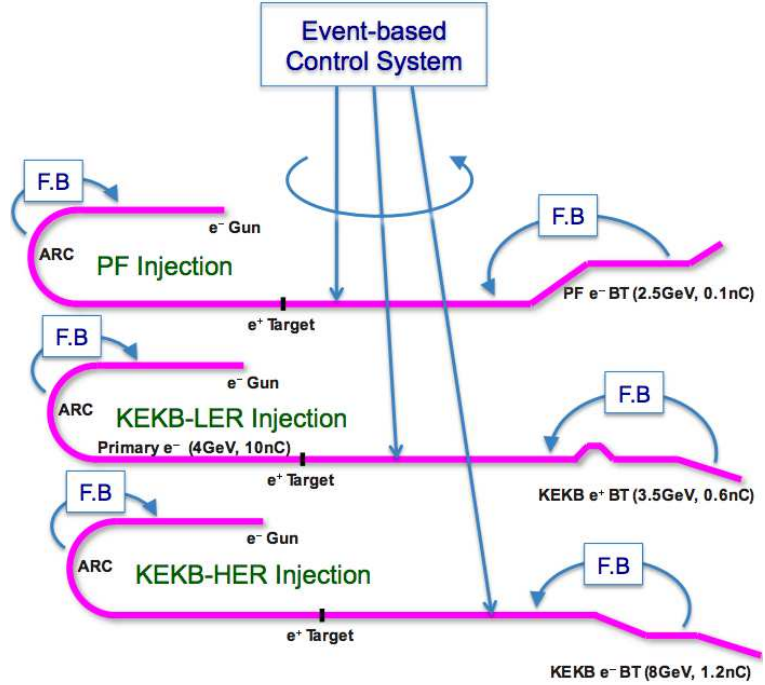}
   \caption{Independent closed feedback loops (F.B.) on three virtual accelerators for KEKB-HER, KEKB-LER, and PF.}
   \label{fig-virt}
\end{figure}

BPM information and RF control parameters were also handled independently in each virtual accelerator.  At first, energy feedback loops at the 180 degree arc and at the end of the linac were installed using event control parameters on each virtual accelerator as in Fig.~\ref{fig-virt}.  As the parameters are independently managed, no modifications to the existing software were necessary.  

The performance of these closed loops was observed with small feedback gains during normal operation.  In these feedback operations no beam stability improvements were achieved.  In other words, no signs of instabilities were observed other than white noise, since the hardware stability was much improved in the later years of the project~\cite{ipac10-ppm}.  

For energy stabilization at the PF beam-transport line, it turned out that the separation of the betatron and dispersion functions was not optimal and the resolution of BPMs was insufficient because of the low beam charges.  The procedures of betatron oscillation compensation and the weighed average of beam positions were applied later.  Because the processing speed with a scripting language was not sufficient in a certain case, the EPICS EPID (enhanced proportional, integral, and derivative controller) record mechanism was employed as well. 

The orbit and energy spread stabilizations can be implemented in the same way.  These beam feedback signals provided valuable information for the accelerator operation.  

We could insert a test beam pulse between injection beam pulses and associate it with one of the virtual accelerators.  Using such a virtual accelerator we could perform a kind of beam study during normal operation.  The same could be applied to one of the stored beam bunches in the ring as well.  This virtual accelerator concept might play an important role in the future SuperKEKB operation. 

\section{Conclusion}

The accelerator control system in the KEKB project achieved successful operation of the project, leading satisfactory physics results with the two basic facilities of the EPICS software toolkit and scripting languages.  Later, two additional concepts of ``channel access everywhere'' and ``dual-tier controls'' were introduced for further improvements.  Based on such control foundations the virtual accelerator mechanism was successfully tested; this could be the basis of next-generation accelerators, including SuperKEKB.  

\section*{Acknowledgments}

The authors are grateful to all the staff and operators at KEKB and the linac for their suggestions and continuous encouragement during the project. 
They also thank the developers from Mitsubishi Electric System \& Service Co.~Ltd. and East Japan Institute of Technology Co.~Ltd. for their system development. 


\end{document}